\documentclass[12pt]{article}
\usepackage[T2A]{fontenc}
\usepackage[utf8]{inputenc}
\usepackage[english]{babel}
\usepackage{amssymb}
\usepackage{amsmath}
\usepackage{graphicx}
\usepackage[small,sc,center]{titlesec}
\usepackage{indentfirst}
\usepackage{siunitx}
\usepackage[labelsep=period]{caption}
\usepackage{caption}

\newcommand{\nc}{\newcommand}
\nc{\er}{\eta_\mathrm{R}}
\nc{\mzahb}{M_\mathrm{ZAHB}}
\nc{\mzams}{M_\mathrm{ZAMS}}
\nc{\tev}{t_\mathrm{ev}}
\nc{\teff}{T_\mathrm{eff}}
\nc{\teffb}{T_\mathrm{eff,b}}
\nc{\teffr}{T_\mathrm{eff,r}}
\begin{document}

\begin{center}
\textbf{On period distribution of RR Lyr type variables in the globular cluster M3}

\vskip 3mm
\copyright\quad
\textbf{2019 г. \quad Yu. A. Fadeyev\footnote{E--mail: fadeyev@inasan.ru}}

\textit{Institute of Astronomy, Russian Academy of Sciences,
        Pyatnitskaya ul. 48, Moscow, 119017 Russia} \\

Received March 14, 2019; after revision April 1, 2019;\\ accepted April 1, 2019
\end{center}

\textbf{Abstract} ---
Evolutionary calculations of population II stars with chemical composition of the globular
cluster M3 were carried out under various assumptions about the initial stellar mass
($0.809M_\odot\le M_\mathrm{ZAMS} \le 0.83M_\odot$) and the mass loss rate parameter
in the Reimers formula ($0.45\le\er\le 0.55$).
In general, 30 evolutionary tracks of the horizontal branch stars were computed.
Selected models of evolutionary sequences were used as initial conditions for solution
of the equations of hydrodynamics that describe radial stellar oscillations.
Hydrodynamic models of RR Lyr type stars were computed for the core helium burning stage
as well as for the preceding pre--ZAHB stage.
Analytic relations for the effective temperature of the instability strip edges
as a function of stellar luminosity are obtained.
Theoretical histograms of the period distribution of RR Lyr type variables
were produced for each evolutionary sequence using Monte--Carlo simulations
based on the consistent stellar evolution and nonlinear stellar
pulsation calculations.
A satisfactory agreement with observations (i.e. the greater number of RRab
variables) was found for the evolutionary sequence
$M_\mathrm{ZAMS} = 0.811M_\odot$, $\eta_\mathrm{R}=0.55$
with the number fraction of fundamental mode pulsators $\approx 75\%$.
At the same time the mean period of fundamental mode pulsators
($\langle\Pi\rangle_0=0.79$ day) is substantially greater compared to the
observational estimate of $\langle\Pi\rangle_\mathrm{ab}$.


\newpage
\section*{introduction}

RR Lyr type pulsating variables observed in globular clusters are low--mass
($M\approx 0.6M_\odot$) population II stars on the evolutionary stage of
steady--state thermonuclear core helium burning (Iben, 1974; Caputo, 1998).
In the Hertzsprung--Russel diagram (HRD) the RR Lyr type variables locate
on the horizontal branch within the pulsational instability strip and have
effective temperatures
$6000~\mathrm{K}\lesssim T_\mathrm{eff} \lesssim 7600~\mathrm{K}$
(Stellingwerf, 1984; Bono, Stellingwerf, 1994).
Most of RR Lyr stars are the fundamental mode or the first overtone pulsators
(i.e. variables of RRab and RRc type, respectively).

It has been shown by Oosterhoff (1939) that galactic globular clusters may be
divided into two groups depending on the mean period and the number fractions
of RRab and RRc variables.
The criteria of the globular clusters of the first group are
$\langle\Pi\rangle_\mathrm{RRab} < 0.6$~day and
$\langle\Pi\rangle_\mathrm{RRc} < 0.35$~day
with the relative number of RRc variables $f_\mathrm{RRc} < 30\%$.
On the contrary, in clusters of the second group
$\langle\Pi\rangle_\mathrm{RRab} > 0.6$~day,
$\langle\Pi\rangle_\mathrm{RRc} > 0.35$~day
and $f_\mathrm{RRc} > 30\%$ (van den Bergh, 1957).
The Oosterhoff dichotomy seems to be due to different physical conditions
corresponding to pulsational mode switching (van Albada, Baker, 1973)
but the cause of the dichotomy remains ubclear.

The globular cluster M3 (NGC~5272) is one of the most thoroughly investigated
clusters due to its relatively small distance and insignificant interstellar reddening.
Recent distance estimates range from 4.44 kpc (Watkins, van der Marel, 2017)
to 10.05 kpc (Marconi et al., 2003) and the interstellar reddening is $E=(B-V)=0.013$
(VandenBerg et al., 2016).
In comparison with other globular clusters M3 is conspicuous due to the large
number of RR Lyr pulsating variables.
By now about 274 RR Lyr variables were discovered in this cluster
(Bakos et al., 2000).
The observational period distribution of RR Lyr type variables
(Corwin, Carney, 2001)
undoubtedly indicate that M3 belongs to the first group of Oosterhoff 
classification.

Theoretical analysis of horizontal branch stars is based on methods of population
synthesis using results of stellar evolution calculations and observations of
RR Lyr type variables.
In the framework of this approach the relations between the pulsation period $\Pi$
and fundamental parameters of the star (the mass $M$, the muninosity $L$,
the effective temperature $\teff$, the metal abundance $Z$)
are obtained from the formulae approximating the extensive grids of hydrodynamic models
of RR Lyr stars (Caputo et al., 1998; Marconi et al., 2003; 2015).
However attempts to reproduce the period distribution of RR Lyr stars in
the globular cluster M3 have failed (Catelan, 2004).

The goal of the present study is to consider the theoretical period distribution
of RR Lyr variables in the globular cluster M3 using the consistent
calculations of stellar evolution and nonlinear stellar pulsations.
In this work we employ various assumptions on the initial stellar mass and
the mass loss rate.
The study implies calculation of the evolutionary sequences of horizontal branch stars
and the use of selected evolutionary models as initial conditions for solution
of the equations of radiation hydrodynamics and time--dependent convection
describing radial stellar oscillations.
Consistent evolutionary and nonlinear pulsation calculations allow us to determine
the stellar age when the star crosses the instability strip edge as well as
the age of the pulsational mode swithing.
Finally, for each evolutionary sequence we produce the theoretical period distribution
which is compared with observations.

\section*{evolutionary tracks of horizontal branch stars}

Following the methodology of our preceding work (Fadeyev, 2018) the evolutionary
sequences of horizontal branch stars were computed from the zero--age main sequence
(ZAMS) to core helium exhaustion.
The nuclear network includes 29 nuclides from hydrogen ${}^1\mathrm{H}$ to
aluminium ${}^{27}\mathrm{Al}$ which are coupled by 51 reactions.
The reaction rates were calculated using the JINA Reaclib data
(Cyburt et al., 2010).
The initial fractional mass abundance of helium was assumed to be
$Y_0=0.25$ (Salaris et al., 2004)
with initial metallicity $Z_0=0.001$ (Catelan, 2004).

Convective mixing was treated in the framework of the standard theory
(B\"ohm--Vitense, 1958) with mixing length to pressure scale height ratio
$\alpha_\mathrm{MLT} = \Lambda/H_\mathrm{P} = 2.0$.
In general, we computed 30 evolutionary sequences with initial masses
$0.809M_\odot\le M_\mathrm{ZAMS} \le 0.83M_\odot$
for the mass loss rate parameter $0.45\le \eta_\mathrm{R}\le 0.55$
in the Reimers formula (1975).
All evolutionary calculations were carried out with the program MESA version 10398
(Paxton et al., 2018).

The main difficulty encountered in evolutionary calculations of the horizontal
branch stars on the stage of core helium burning is due to the jump
of the helium abundance on the outer boundary of the convective core.
Finite--difference representation of the evolutionary model is responsible for
discrete ingestion of the helium--rich material into the convective core.
As a result, the energy generation rate in the triple--$\alpha$ reactions may
substantially increase so that the evolution time on the horizontal branch
significantly increases due to spurious loops in the HRD
(Constantino et al., 2015; 2016).
In the present study to avoid appearence of spurious breathing pulses we
employed the method that allows us to constrain the ingestion rate of unburned
helium--rich material on the outer boundary of the convective core
(Spruit, 2015; Constantino et al., 2017).

Selected models of evolutionary sequences located in the vicinity of the
instability strip in the HRD were used as initial conditions for solution
of the equations of radiation hydrodynamics describing radial stellar
oscillations.
Basic equations and the choice of parameters of the time--dependent convection
theory (Kuhfu\ss, 1986) are discussed in our preceding papers
(Fadeyev, 2013; 2015).
In the present study the solution of the transport equations for turbulent
convection was carried out with $\alpha_\mu=0.3$ in the
expression for the turbulent viscosity
\begin{equation}
 \mu = \alpha_\mu \rho \Lambda E_\mathrm{trb}^{1/2} ,
\end{equation}
where $\rho$ is the gas density, $\Lambda$ is the mean free path of the turbulent element
(i.e. the mixing length), $E_\mathrm{trb}$ is the mean kinetic energy of turbulence.
The parameter $\alpha_\mu$ determines efficiency of ineraction between turbulent
elements and the gas flow.
For pulsating stars this parameter ranges within $0.1 < \alpha_\mu < 0.5$
and the pulsation period is almost independent of $\alpha_\mu$
(Wuchterl, Feuchtinger, 1998; Olivier, Wood, 2005; Smolec, Moskalik, 2008).

For each crossing of the instability strip we computed from 10 to 15 hydrodynamic models.
The edge of the instability strip was determined using two adjacent models
one of which is unstable against radial pulsations and another shows decaying oscillations.
The growth (or decay) rate of oscillations is $\eta=\Pi^{-1} d\ln E_\mathrm{K,max}/dt$,
where $\Pi$ is the period of radial stellar pulsations,
$E_\mathrm{K,max}$ is the maximum kinetic energy of pulsation motions.
The kinetic energy reaches its maximum value twice per pulsation period.
The evolutionary time $\tev$ (i.e. the star age) correspoinding to the instability edge
($\eta=0$) was evaluated by linear interpolation of $\eta(\tev)$.

Effective temperatures of the blue $\teffb$ and the red $\teffr$ edges of the instability
strip can be expressed as a function of the bolometric stellar luminosity $L$ by
following relations
\begin{gather}
\label{blue}
\log\teffb = 4.0561 - 0.1040 \log (L/L_\odot) ,
\\
\label{red}
\log\teffr = 3.9847 - 0.1096 \log (L/L_\odot) .
\end{gather}
The constant coefficients in (\ref{blue}) and (\ref{red}) were obtained by
the least--squares method within $1.638 < \log L/L_\odot < 1.883$
for 58 and 84 crossings of the blue and red edges of the instability strip,
respectively, by 30 evolutionary tracks.
The larger number of red edge crossings is due to decaying oscillations
of energy generation in the triple--$\alpha$ reactions after the helium flash
on the tip of the red giant branch.
Formulae (\ref{blue}) and (\ref{red}) give the effective temperature
on the instability edge with mean r.m.s. deviations of
$\sigma(\log\teffb) \approx 10^{-3}$ and $\sigma(\log\teffr) \approx 1.2\times 10^{-3}$.
The scatter of $\eta=0$ points around the regression line is mainly due to
the linear intrerpolation errors.
The average width of the instability strip is $\Delta\log\teff = 0.081$.

Location of the evolutionary track in the HRD relative to the instability strip
depends on the initial stellar mass $\mzams$ and the mass lost
during the preceding stage of the red giant (i.e. the parameter $\er$).
Dependence of the evolutionary track on the initial mass is illustrated in Fig.~\ref{fig1}
for evolutionary sequences $\mzams=0.81M_\odot$ and $\mzams=0.83M_\odot$
computed for the mass loss rate parameter $\er=0.5$.
The role of mass loss is illustrated in Fig.~\ref{fig2} for evolutionary sequences
$\mzams=0.82M_\odot$ computed for $\er=0.45$, 0.5 and 0.55.

Solid lines in Figs.~\ref{fig1} and \ref{fig2} correspond to the evolutionary stage
of core helium burning.
For all evolutionary sequences considered in the present study the time of
core helium burning with an accuracy of a few percent is $t_\mathrm{HB}\approx 10^8$~yr.
The preceding evolutionary stage when the star leaves the tip of the red giant branch
and approaches the horizontal branch is significantly shorter.
For example, the time interval between the maximum energy generation during the helium
flash and commencement of core helium burning is $\approx 1.4\times 10^6$~yr.
In Figs.~\ref{fig1} and \ref{fig2} this stage of evolution is shown by dotted lines
and in close vicinity of the instability strip its duration is
$t_\textrm{pre-ZAHB}\lesssim 10^6$~yr.
The point between the solid and dotted lines corresponds to the horizontal branch
of the zero age (ZAHB).

\section*{periods of radial pulsations}

Computations of each hydrodynamic model were carried out on the time interval encompassing
hundreds of pulsation cycles.
After completion of hydrodynamic computations the pulsation periods of the fundamental
mode $\Pi_0$ and the first overtone $\Pi_1$ were evaluated using the discrete Fourier
transform of the pulsating stellar envelope kinetic energy.
Both periods are time--independent because after attainment of the limiting amplitude
nonlinear effects remain negligible.
Therefore, to enhance the accuracy of period determination we calculated the power
spectrum of the kinetic energy for the whole time interval of the solution
of the equations of hydrodynamics.
While approaching the limiting cycle the amplitude of one mode in the spectrum reduces
whereas another mode becomes prevalent.

During the instability strip crossing the star undergoes $\sim 10^9$ oscillations
on the stage of core helium burning and $\sim 10^7$ oscillations on the
pre--ZAHB stage.
Therefore, we may assume with high accuracy that the mode switching occurs instantly.
In the present study the stellar age $\tev$ corresponding to the mode switching was determined
as a mean evolutionary time of two adjacent hydrodynamic models pulsating in different modes.

The pulsation period as a continuous function of evolutionary time $\Pi(\tev)$
was determined using the cubic interpolation splines.
Results of approximation are shown in Figs.~\ref{fig3} and \ref{fig4} for
evolutionary sequences displayed in Figs.~\ref{fig1} and \ref{fig2}.
For the sake of graphical clarity the evolutionary time $\tev$ is set to zero at ZAHB,
the time scale for $\tev < 0$ being nearly two orders of magnitude shorter in
comparison with that for $\tev > 0$.
Each plot in Figs.~\ref{fig3} and \ref{fig4} represents the temporal dependence of the
pulsation period within the instability strip and a discontinuous jump
corresponds to the mode switching.

As clearly seen in Figs.~\ref{fig3} and \ref{fig4} the lifetime of the star within
the instability strip $t_\mathrm{RR}$
significantly reduces with decreasing mass of the horizontal branch
star (i.e. with decreasing $\mzams$ or increasing $\er$).
Approximate estimates of $t_\mathrm{RR}$ in units of the core helium burning
time $t_\mathrm{HB}$ are given in the second column of the Table~\ref{table1}
for several values of the zero age horizontal branch stellar mass $\mzahb$.
The instability strip lifitime on the pre--ZAHB stage $t_\mathrm{RR,pre-ZAHB}$
is almost independent of stellar mass,
so that decrease of $\mzahb$ is accompanied by increasing number fraction of
pre--ZAHB RR~Lyr variables (see the third column in Table~\ref{table1}).

\section*{period distribution}

Theoretical histograms of RR~Lyr period distribution were produced for all
computed evolutionary sequences using $10^7$ Monte--Carlo simulations.
To this end the age of the synthetic star $\tev$ was randomly selected from
a uniform distribution on the interval $t_\textrm{pre-ZAHB}\le\tev\le t_\mathrm{HB}$,
where $t_\textrm{pre-ZAHB}=-10^6$~yr.

First of all, we have to note that histograms of evolutionary sequences
computed with $\er\le 0.5$ should be excluded from our consideration
due to contradiction with observations because
they exhibit significant excess of the first overtone pulsators.
Moreover, as clearly seen in Fig.~\ref{fig2}, the maximum effective
temperature of the horizontal branch stars seems to be insufficiently
high in comparison with its observational estimates
(Catelan et al., 2001; Cacciari et al., 2005).

Fig.~\ref{fig5} shows the normalized histograms of the period distribution
produced from evolutionary and stellar pulsation computations of three
evolutionary sequences with initial masses
$\mzams=0.81M_\odot$, $0.818M_\odot$, $0.828$ and the mass loss rate parameter $\er=0.55$.
The masses of RR~Lyr stars are
$M=0.578M_\odot$, $0.589M_\odot$ and $0.603M_\odot$, respectively.
As clearly seen, to avoid the contradiction with observations and
to obtain excess of fundamental mode pulsators we have to assume that
the initial mass is $\mzams=0.81M_\odot$.

For a more detailed comparison between results of computations and observations
Fig.~\ref{fig6} shows normalized histograms of evolutionary sequences with
initial masses $0.809M_\odot\le\mzams\le 0.812M_\odot$ where each plot
is accompanied by the number fraction of fundamental mode pulsators $f_0$.
As clearly seen, the satisfactory agreement with observations is obtained
for evolutionary sequences $\mzams=0.810M_\odot$ and $\mzams=0.811M_\odot$,
that is for RR Lyr stars with masses $0.578M_\odot\le M\le 0.580M_\odot$.
The number fractions of fundamental mode pulsators (71\% and 75\%)
are enough close to 78\% of RR~Lyr variables in the globular cluster M3
(Catelan, 2004).
At the same time one should note that the mean period of fundamental
mode pulsators $\langle\Pi\rangle_0=0.79$ day is significantly larger
in comparison with the mean period of RRab variables
$\langle\Pi\rangle_\mathrm{ab}=0.56$ day.

The existence of the narrow interval of intial masses $\mzams$ with satisfactory
agreement between theory and observations is due to the fact that
the mode switching is sensitively dependent on the stellar mass.
This dependence is illustrated in Fig.~\ref{fig7} where the periods of the
fundamental mode and the first overtone at the mode switching
are plotted as a function of $\mzams$.
The plots correspond to the final core helium burning stage when radial
pulsation mode changes from the first overtone to the fundamental mode.
As is seen, in the initial mass interval $0.809\le\mzams\le 0.812M_\odot$
the period of each mode at the mode switching varies by $\approx 15\%$.
A strong sensitivity of the pulsation period at the mode switching
to the stellar mass is the main cause of different shapes of histograms
displayed in Fig.~\ref{fig6}.

\section*{conclusions}

In the present study we carried out consistent calculations of stellar evolution
and nonlinear stellar pulsations for 30 evolutionary sequences of horizontal
branch stars.
Initial masses of evolutionary models correspond to the age of RR~Lyr stars
from $1.195\times 10^{10}$ yr for $\mzams=0.83M_\odot$
to $1.309\times 10^{10}$ yr for $\mzams=0.809M_\odot$.
Almost all theoretical histograms exhibit excess of short--period RRc type
variables and contradict to observations of the globular cluster M3.
A satisfactory agreement with observations was obtained for two evolutionary
sequences $\mzams=0.810M_\odot$ and $0.811M_\odot$ computed with the mass loss
rate parameter $\er=0.55$.
The number fraction of fundamental mode pulsators (71\% and 75\%) was found to be
insignificantly smaller than the observed number fraction of RRab
variables (78\%).

The evolutionary sequences of horizontal branch stars computed in the present study
will be used to improve the theoretical estimate of the mean period of RR Lyr
variables pulsating in the fundamental mode.
Bearing in mind the fact that the mode switching is strongly sensitive to the
stellar mass we should consider in the future work the evolutionary sequences
computed in vicinity of the values $\mzams=0.811M_\odot$, $\er=0.55$ and $Y_0=0.25$.

Solution of the problem is complicated by requirements of more thorough evolutionary
calculations.
The instability strip crossing on the stage of core helium burning occurs when
the fractional mass abundance of helium in the convective core is $Y < 0.1$,
so that effects of unsteady mass ingestion on the outer boundary of the convective
core become stronger and lead to perceptible changes in evolutionary calculations.
As was shown above reduction of such effects is necessary for correct
determination of the mode switching and thereby for better agreement
between theory and observations.

The author is indebted to L.R. Yungelson for critical comments and
useful discussions.

\newpage
\section*{references}

\begin{enumerate}

\item G.A. Bakos, J.M. Benko, and J. Jurcsik, Acta Astron. \textbf{50}, 221 (2000).

\item E. B\"ohm--Vitense, Zeitschrift f\"ur Astrophys. \textbf{46}, 108 (1958).

\item G. Bono and R.F. Stellingwerf, Astrophys. J. Suppl. Ser. \textbf{93}, 233 (1994).

\item C. Cacciari, T.M. Corwin, and B.W. Carney, Astron. J. \textbf{129}, 267 (2005).

\item F. Caputo, Astron. Astrohys. Rev. \textbf{9}, 33 (1998).

\item F. Caputo, P. Santolamazza, and M. Marconi, MNRAS \textbf{293}, 364 (1998).

\item M. Catelan, Astrophys. J. \textbf{600}, 409 (2004).

\item M. Catelan, F.R. Ferraro, and R.T. Rood, Astrophys. J. \textbf{560}, 970 (2001).

\item T. Constantino, S.W. Campbell, J. Christensen--Dalsgaard, J.C. Lattanzio, and D. Stello, MNRAS \textbf{452}, 123 (2015).

\item T. Constantino, S.W. Campbell, W. Simon, J,C. Lattanzio, and A. van Duijneveldt, MNRAS, \textbf{456}, 3866 (2016).

\item T. Constantino, S.W. Campbell, and J.C. Lattanzio, MNRAS \textbf{472}, 4900 (2017).

\item T.M. Corwin and B.W. Carney, Astron. J. \textbf{122}, 3183 (2001).

\item R.H. Cyburt, A.M. Amthor, R. Ferguson, Z. Meisel, K. Smith,
      S. Warren, A. Heger, R.D. Hoffman, T. Rauscher, A. Sakharuk, H. Schatz, 
      F.K. Thielemann, and M. Wiescher, Astrophys. J. Suppl. Ser. \textbf{189}, 240 (2010).

\item Yu.A. Fadeyev, Astron. Lett. \textbf{39}, 306 (2013).

\item Yu.A. Fadeyev, MNRAS \textbf{449}, 1011 (2015).

\item Yu.A. Fadeyev, Astron. Lett. \textbf{44}, 616 (2018).

\item I. Iben, Ann. Rev. Astron. Astrophys. \textbf{12}, 215 (1974).

\item R. Kuhfu\ss, Astron. Astrophys. \textbf{160}, 116 (1986).

\item M. Marconi, F. Caputo, M. Di Criscienzo, and M. Castellani, Astrophys. J. \textbf{596}, 299 (2003).

\item M. Marconi, G. Coppola, G. Bono, V. Braga, A. Pietrinferni, R. Buonanno,M. Castellani,
      I. Musella, V. Ripepi, and R.F. Stellingwerf), Astrophys. J. \textbf{808}, 50 (2015).

\item E.A. Olivier and P.R. Wood, MNRAS \textbf{362}, 1396 (2005).

\item P.T. Oosterhoff, Observatory \textbf{62}, 104 (1939).

\item B. Paxton, J. Schwab,  E.B. Bauer, L. Bildsten, S. Blinnikov, P. Duffell,
      R. Farmer, J.A. Goldberg, et al., Astropys. J. Suppl. Ser. \textbf{234}, 34 (2018).

\item D. Reimers, \textit{Problems in stellar atmospheres and envelopes}
      (Ed. B. Baschek, W.H. Kegel, G. Traving, New York: Springer-Verlag, 1975), p. 229.

\item M. Salaris, M. Riello, S. Cassisi, and G. Piotto, Astron. Astrophys. \textbf{420}, 911 (2004).

\item R. Smolec and P. Moskalik, Acta Astron. \textbf{58}, 193 (2008).

\item H.C. Spruit, Astron. Astrophys. \textbf{582}, L2 (2015).

\item R.F. Stellingwerf, Astrophys. J. \textbf{277}, 322 (1984).

\item D.A. VandenBerg, P.A. Denissenkov, and M. Catelan, Astrophys. J. \textbf{827}, 2 (2016).

\item T.S. van Albada and N. Baker, Astrophys. J. \textbf{185}, 477 1973.

\item S. van den Bergh, 1957, Astron. J. \textbf{62}, 334 (1957).

\item L.L. Watkins and R.P. van der Marel, Astrophys. J. \textbf{839}, 89 (2017).

\item G. Wuchterl and M.U. Feuchtinger, Astron. Astrophys. \textbf{340}, 419 (1998).

\end{enumerate}

\newpage
\begin{table}
\caption{Characteristic lifetimes of RR~Lyr stars}
\label{table1}
\begin{center}
 \begin{tabular}{clc}
  \hline
  $\mzahb/M_\odot$ \quad & $t_\mathrm{RR}/t_\mathrm{HB}$ & $t_\mathrm{RR,pre-ZAHB}/t_\mathrm{RR}$ \\
  \hline
      0.58  & 0.003 & 0.42 \\
      0.59  & 0.004 & 0.10 \\
      0.60  & 0.01  & 0.12 \\
      0.61  & 0.07  & 0.04 \\
  \hline          
 \end{tabular}
\end{center}
\end{table}
\clearpage

\newpage
\section*{Figure captions}

\begin{figure}
\centerline{\includegraphics[width=14cm]{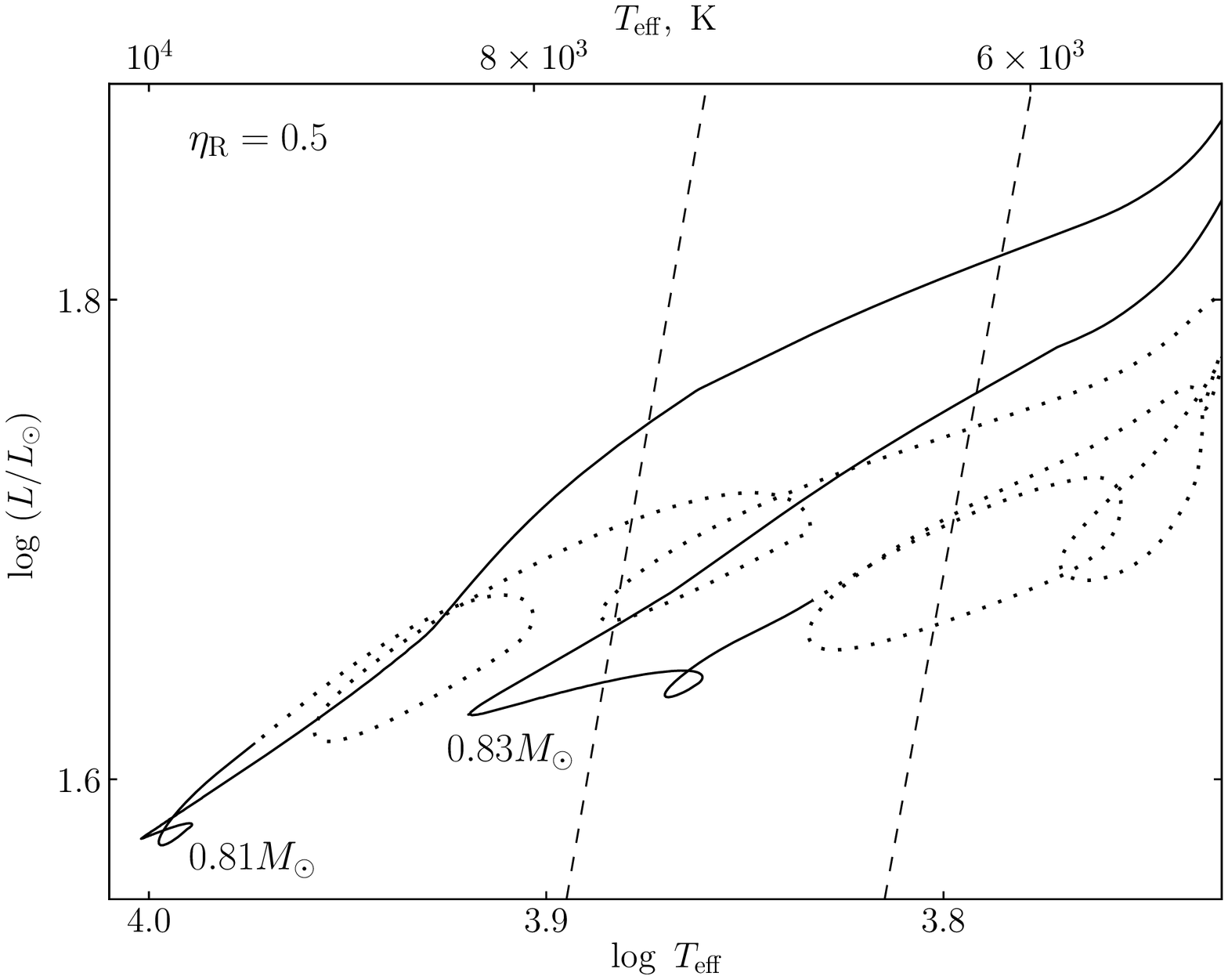}}
\caption{}
\label{fig1}
\end{figure}

\begin{figure}
\centerline{\includegraphics[width=14cm]{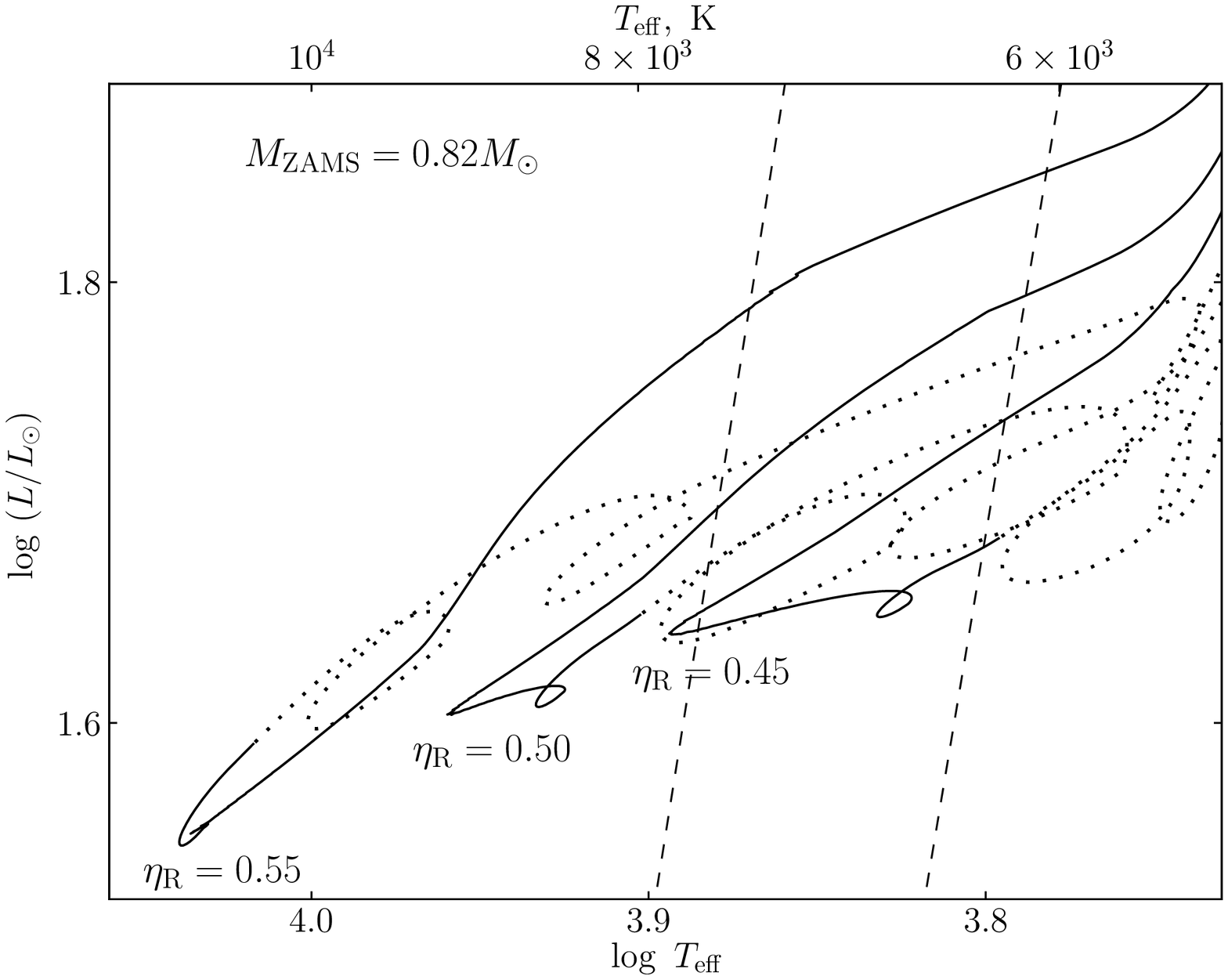}}
\caption{}
\label{fig2}
\end{figure}

\begin{figure}
\centerline{\includegraphics[width=14cm]{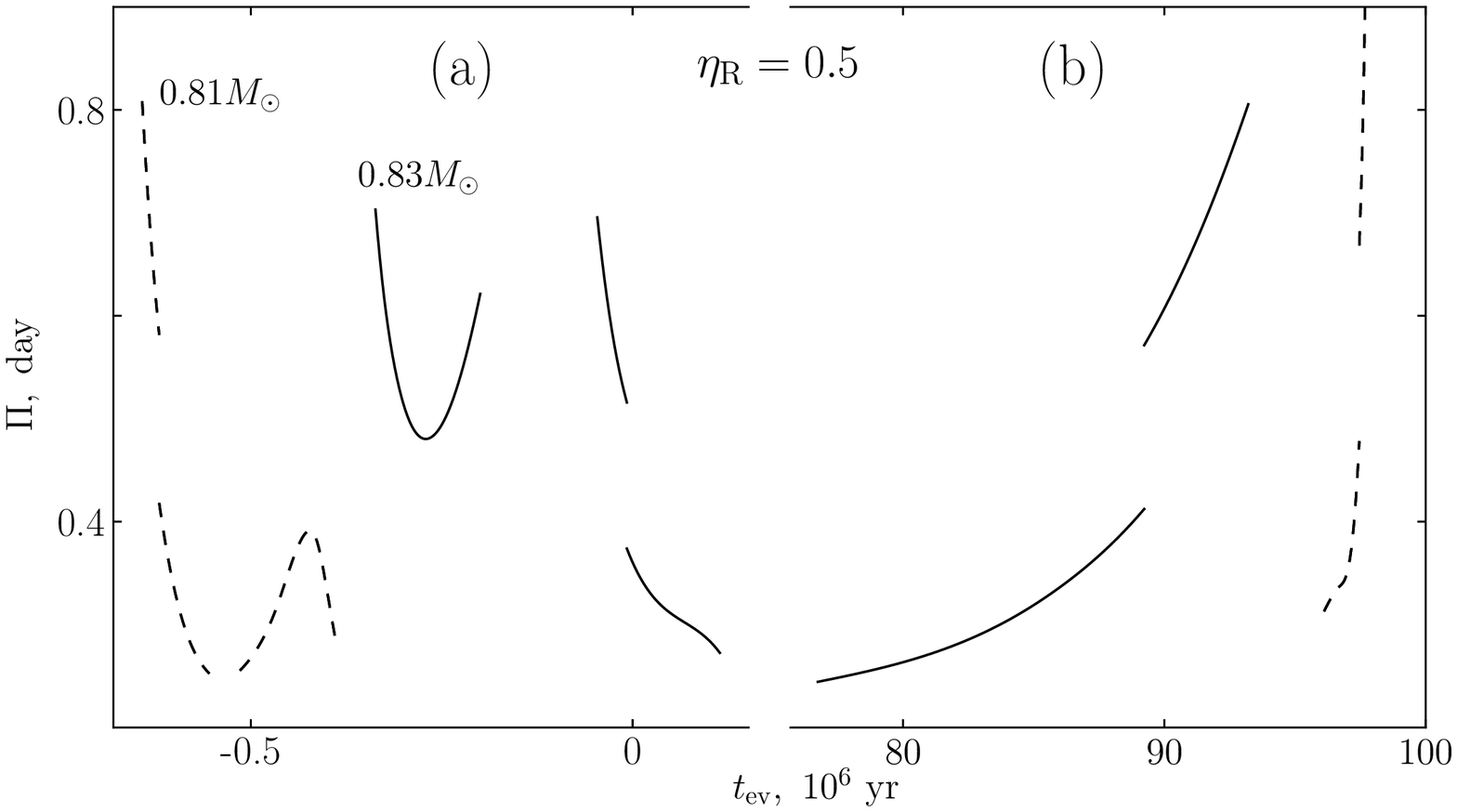}}
\caption{}
\label{fig3}
\end{figure}

\begin{figure}
\centerline{\includegraphics[width=14cm]{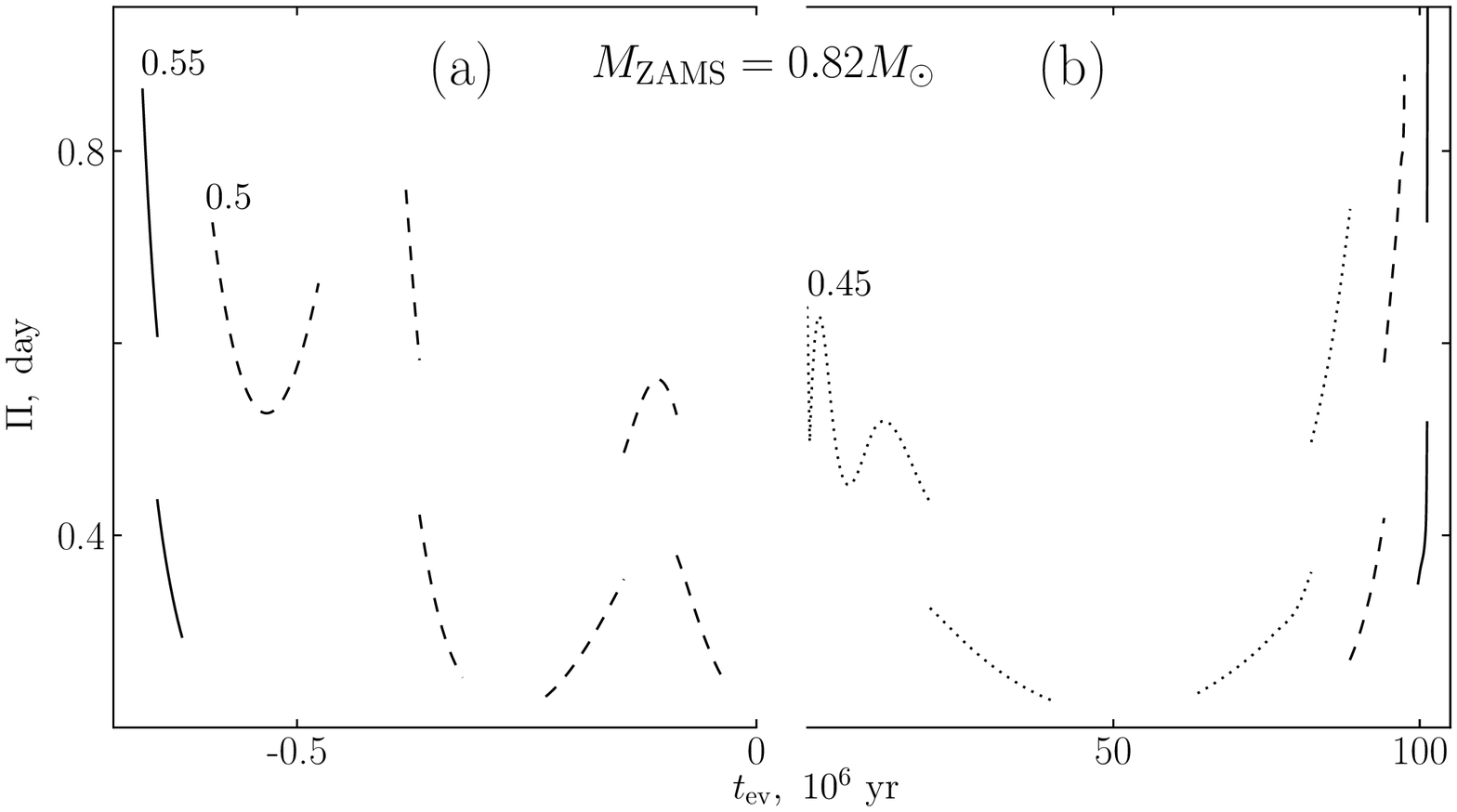}}
\caption{}
\label{fig4}
\end{figure}

\begin{figure}
\centerline{\includegraphics[width=18cm]{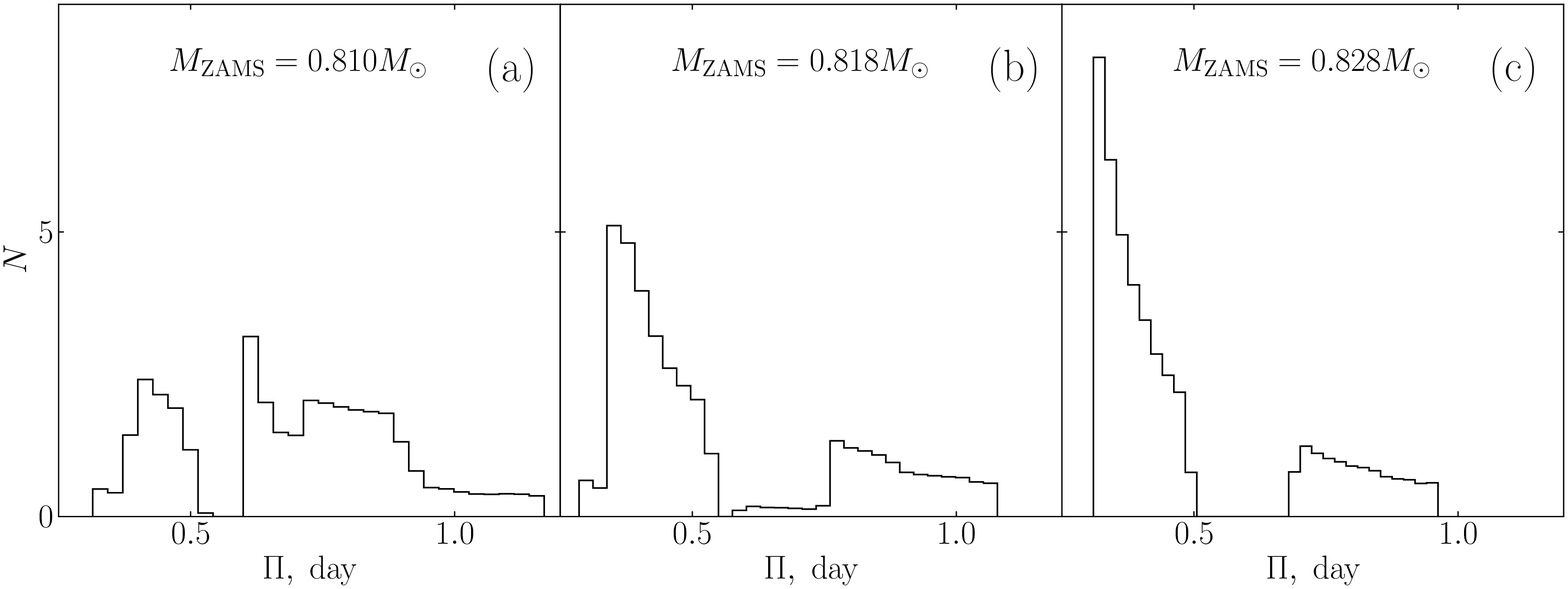}}
\caption{}
\label{fig5}
\end{figure}

\begin{figure}
\centerline{\includegraphics[width=18cm]{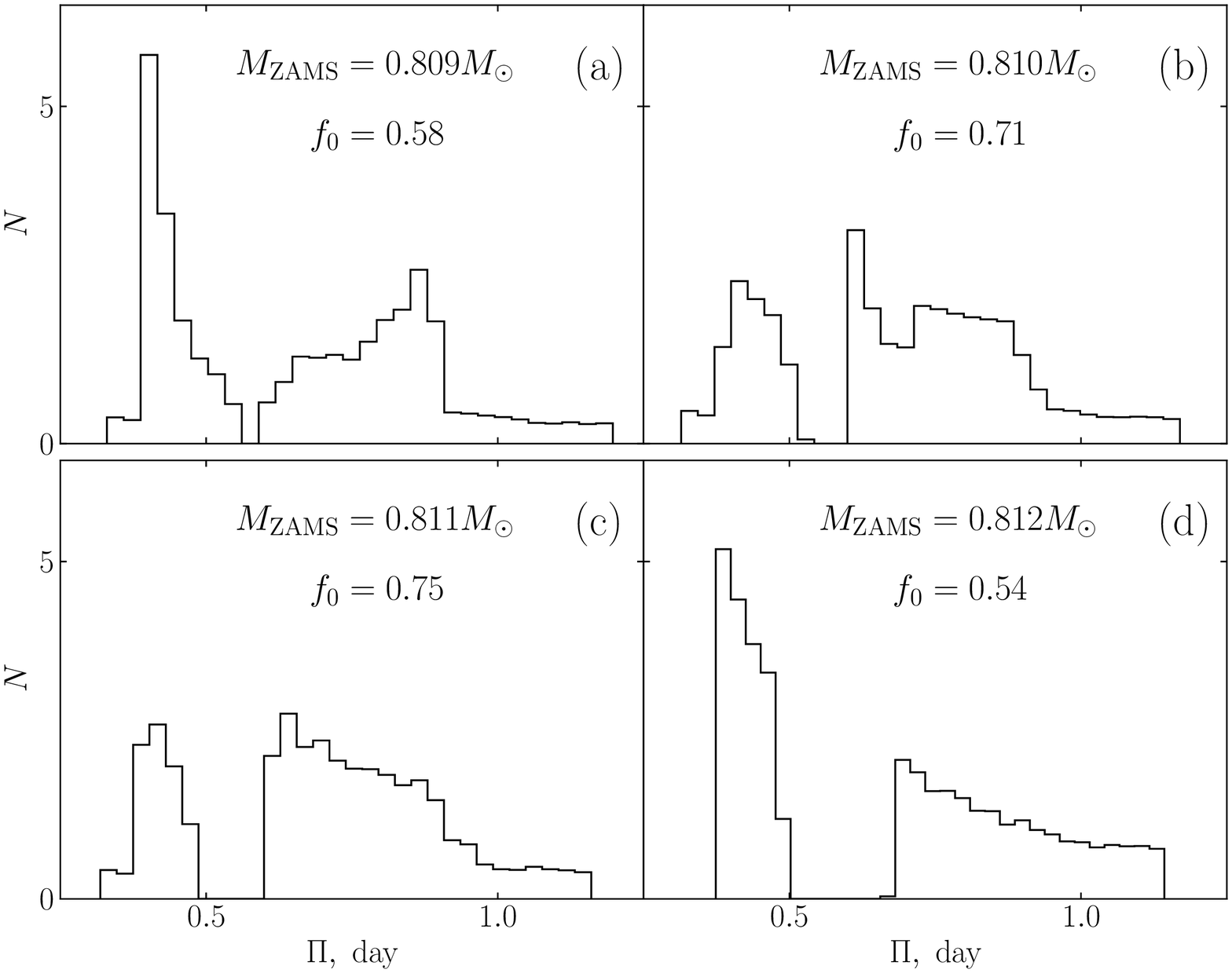}}
\caption{}
\label{fig6}
\end{figure}

\begin{figure}
\centerline{\includegraphics[width=16cm]{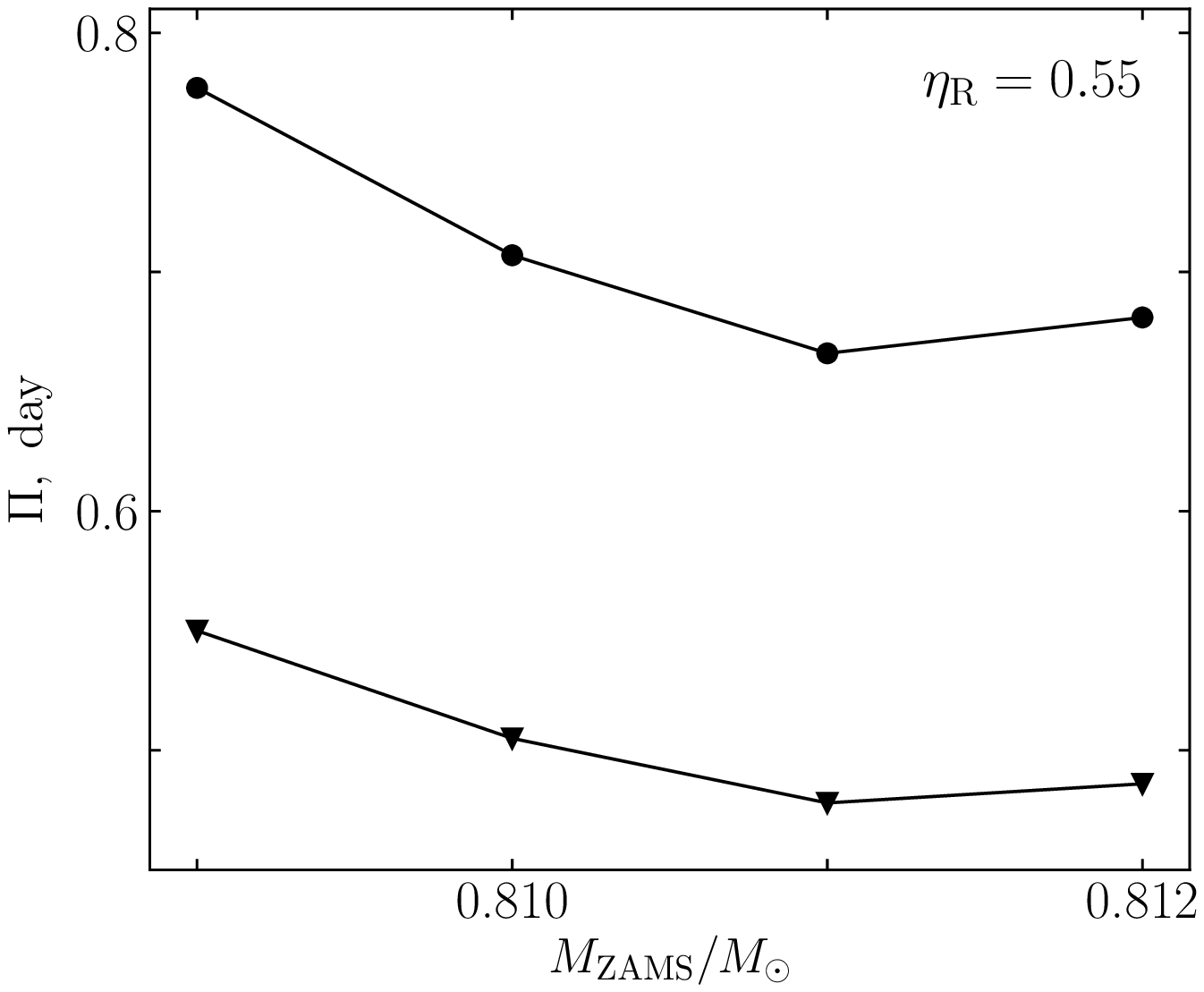}}
\caption{}
\label{fig7}
\end{figure}

\begin{itemize}
 \item[Fig. 1.] {Evolutionary tracks of horizontal branch stars with initial masses
         $\mzams=0.81M_\odot$ and $\mzams=0.83M_\odot$ for the mass loss parameter $\er=0.5$.
         The stage of core helium burning and the preceding pre--ZAHB stage
         are shown by solid and dotted lines, respectively.
         Dashed lines indicate the edges of the instability strip given
         by relations (\ref{blue}) и (\ref{red}).}

\item[Fig. 2.] {Same as Fig.~\ref{fig1} but for evolutionary sequences $\mzams=0.82M_\odot$
         computed with mass loss rate parameter $\er=0.45$, 0.5 and 0.55.}

\item[Fig. 3.] {The pulsation period $\Pi$ as a function of stellar age $\tev$ for
         evolutionary sequences with initial masses $\mzams=0.81M_\odot$ (dashed lines) and
         $\mzams=0.83M_\odot$ (solid lines) computed for the mass loss rate parameter $\er=0.5$.
         The stellar age is set to zero at ZAHB.}

\item[Fig. 4.] {Same as Fig.~\ref{fig3} but for evolutionary sequences $\mzams=0.82M_\odot$
         computed for the mass loss rate parameters $\er=0.45$ (dotted lines),
         0.5 (dashed lines) and 0.55 (solid lines).}

\item[Fig. 5.] {Normalized period distribution histograms of RR Lyr variables
         for evolutionary sequences
         $\mzams=0.81 M_\odot$ (a), $\mzams=0.818M_\odot$ (b) and
         $\mzams=0.828M_\odot$ (c) computed for the mass loss rate parameter $\er=0.55$.}

\item[Fig. 6.] {Same as Fig.~\ref{fig5} but for evolutionary sequences
         $\mzams=0.809 M_\odot$ (a), $\mzams=0.810 M_\odot$ (b),
         $\mzams=0.811 M_\odot$ (c) and $\mzams=0.812M_\odot$ (d).
         $f_0$ is the number fraction of fundamental mode pulsators.}

\item[Fig. 7.] {The pulsation period $\Pi$ of the fundamental mode (filled circles) and
         the first overtone (filled triangles) at the mode switching
         on the stage of core helium burning
         as a function of initial mass $\mzams$ for the mass loss parameter $\er=0.55$.}

\end{itemize}

\newpage
\clearpage

\newpage
\clearpage

\newpage
\clearpage

\newpage
\clearpage

\newpage
\clearpage

\newpage
\clearpage

\newpage

\end{document}